\documentstyle[preprint,aps,prl,epsf]{revtex}
\input{epsf}
\input epsf.sty
\begin{document}
\draft
\title{
Topological Excitations in Double-Layer Quantum Hall systems
}
\author{Kyungsun Moon}
\address{
Department of Physics and Astronomy, University of Oklahoma, Norman,
OK~~73019\\
}
\date{\today}
\maketitle

{\tightenlines
\begin{abstract}
Double-layer quantum Hall systems with spontaneous
broken symmetry can exhibit a novel manybody quantum Hall effect
due to the strong interlayer coherence.
When the layer separation becomes close to the critical value,
quantum fluctuations can destroy the interlayer coherence and
the quantum Hall effect will disappear.
We calculate the renormalized isospin stiffness $\rho_s$ 
due to quantum fluctuations within the Hartree-Fock-RPA formalism. 
The activation energy of the topological excitations thus obtained  
demonstrates a nice qualitative agreement with recent experiment.
\end{abstract}
}

%IF ONLY THREE PACS ALLOWED, PLASE DROP THE LAST ONE

\narrowtext

Double-layer quantum Hall systems (DLQHS) have drawn much theoretical 
and experimental attention recently 
\cite{GirvinMacdBook,DBPRL,Sheena,WenandZee,Ezawa,DBPRBI,Fertig,Shayegan}.
When the distance $d$ between the layers is comparable to the mean particle 
spacing,   
strong interlayer correlations induce a novel manybody quantum Hall effect 
at $\nu_T=1/m$ ($m$ is an odd integer), where $\nu_T$ is
the total filling factor
\cite{Sheena,Shayegan,Hyndman}.
Since each layer at $\nu=1/(2m)$ alone does not support quantum Hall effect,
it manifests a {\em genuine} manybody correlation effect.
When the layer separation becomes large and close to the
critical value, 
quantum fluctuations can destroy the interlayer coherence and
accordingly quantum Hall effect will disappear
\cite{Sheena,Hyndman,MacdPlatBoe}.
By mapping the layer degrees of freedom onto the $S={1\over 2}$ 
isospin variables,
the interlayer correlations can be quantified as isospin stiffness $\rho_s$.
Since most of the recent experimental systems\cite{Sheena,Hyndman} 
belong to the quantum 
fluctuation dominated regime, it is {\em crucial} to obtain the
reliable estimate of the renormalized isospin stiffness $\rho_s$ 
which incorporates the effect of quantum fluctuations.

In this Letter, we calculate the renormalized isospin stiffness $\rho_s$ and 
the effective tunneling amplitude $t_R$ 
within the Hartree-Fock-RPA formalism.
By taking into account the important vertex 
corrections\cite{Fertig,MacdPlatBoe,FertigDB}, we explicitly show 
that $\rho_s$ vanishes 
at $d/\ell\cong 1$, where $\ell=\sqrt{\hbar c/|e|B}$ is 
the magnetic length.
The activation energy of the lowest charge 
excitations ({\em e.g.} Meron pairs {\em or} domain-wall string solitons) is  
estimated and shows a nice qualitative agreement with the recent 
experiment\cite{Hyndman}.
We consider the DLQHS with finite tunneling $t$ and
parallel magnetic field $B_{||}{\hat y}$.
We concentrate on the $\nu_T=1$ quantum Hall effect and 
assume that the electrons are lying on the lowest
Landau level due to the strong perpendicular magnetic 
field $B_\perp {\hat z}$
and the real spins are fully spin-polarized due to the finite Zeeman gap.
With the choice of the following 
gauge ${\bf A}=(0, B_\perp x, -B_\parallel x)$,  
the Hamiltonian of the system can be written by  
\begin{eqnarray}
{\cal H}&=&-t e^{-Q^2/4}\sum_X \left( e^{iQX} c_{X\uparrow}^\dagger 
c_{X\downarrow} 
+ e^{-iQX} c_{X\downarrow}^\dagger c_{X\uparrow} \right) \nonumber\\
&+& {1\over 2}\sum_{\sigma,\sigma'} \sum_{X_1,X_2,X' \ne 0}
V_{\sigma \sigma'}(X',X_1-X_2) c^\dagger_{X_1-X' \sigma}
c^\dagger_{X_2 \sigma'} c_{X_2-X' \sigma'} c_{X_1 \sigma}
\end{eqnarray}
where $\sigma= +1(-1)$ represents an electron in the upper(lower) layer, 
$X$ the guiding center coordinate, and $Q=dB_{||}/\ell^2 B_\perp$.
With the following convention of the units $e=\ell=\hbar=1$, 
$V_{\sigma \sigma'}$ is an intralayer interaction $V_A$ for 
$\sigma=\sigma'$ and an interlayer interaction $V_E$ for $\sigma\ne \sigma'$:
${\tilde V}_A=2\pi/\epsilon q$, ${\tilde V}_E=2\pi e^{-q d}/\epsilon q$, and   
\begin{equation}
V_{\sigma \sigma'}(X,X_1-X_2)=\frac {1}{L^2} \sum_{q_y} e^{-q^2/2} 
\tilde {V}_{\sigma \sigma'}(X,q_y) e^{iq_y(X_1-X_2)}
\end{equation}
where $q=\sqrt{X^2+q_y^2}$.  
We notice that $B_{||}$ induces an Aharonov-Bohm phase
$e^{\pm iQX}$ depending on the sense of interlayer tunneling.

The relatively weak interlayer interaction comparing to the intralayer
interaction supresses local density 
fluctuations. Since this corresponds to the $m_z^2$-term 
in the magnetic analogy, 
the isospins are forced to lie on the ${\hat x}-{\hat y}$ 
isospin space\cite{DBPRBI}.
Hence the DLQHS can be viewed as quantum 
$XY$-ferromagnets.
We obtain the following energy functional  
\begin{equation}
E[\theta]=\int d^2r~\Biggl\{ \frac{1}{2}~\rho_s ({\bf\nabla}\theta)^2 -
{t\over 2\pi}~\cos{[\theta ({\bf r}) - Qx]}\Biggr\} 
\end{equation}
where $\theta({\bf r})$ represents the isospin orientation.
This is precisely the Pokrovsky-Talapov (PT) model\cite{bak}, which 
exhibits a commensurate-incommensurate transition.
For small $Q$ and hence $B_{\parallel}$, the phase obeys
$\theta({\bf r})= Qx$. As $B_{\parallel}$ increases, 
the local field tumbles too rapidly and a continuous
phase transition to an incommensurate state with broken
translation symmetry occurs\cite{DBPRL,DBPRBI}.  
The ground state energy of the commensurate state is given by 
\begin{equation}
\epsilon(Q,t)\equiv \frac {E[Q,t]}{A}={1\over 2}~\rho_s Q^2 
- \frac {t}{2\pi} e^{-Q^2/4}
\end{equation}
where $\rho_s$ measures the interlayer coherence.
By imposing the following condition $\sqrt{t/\rho_s}\gg Q > 0$, 
one can guarantee that 
the ground state of the system is the commensurate one and 
$\rho_s$ can be obtained by the following procedure  
\begin{equation}
\rho_s = \lim_{t\rightarrow 0}\lim_{Q\rightarrow 0} 
\frac {d^2\epsilon (Q,t)}{dQ^2} .
\label{eq:stiff}
\end{equation}
An appealing feature of the commensurate state is that the
self-consistent Hartree-Fock ground state
is also an eigenstate of the tunneling Hamiltonian alone.
One can easily diagonalize the tunneling Hamiltonian in terms of 
the $\alpha_X$,$\beta_X$-operators 
\begin{equation}
H_t=-t e^{-Q^2/4}\sum_X (\alpha_X^\dagger \alpha_X
-\beta_X^\dagger \beta_X)
\end{equation}
where $\alpha_X,\beta_X$ 
correspond to the symmetric and antisymmetric state with respect to
the spatially varying tumbling field:
$\alpha_X =(e^{-iQ X/2} c_{X\uparrow}+e^{iQ X/2}
c_{X\downarrow})/\sqrt{2}$, $\beta_X=(e^{-iQ X/2} c_{X\uparrow}-e^{iQ X/2}
c_{X\downarrow})/\sqrt{2}$.
The ground state $|\Phi\rangle$ is a completely filled state 
of the $\alpha_X$-particles. 
Since $|\Phi\rangle$ is an eigenstate of the full Hartree-Fock 
Hamiltonian {\em too}, 
it is straightforward to calculate the Hartree-Fock ground state energy
yielding 
\begin{eqnarray}
\frac {E_{HF}}{A}&=&\frac {1}{A}\langle\Phi|{\cal H}|
\Phi\rangle\nonumber\\
&=&-{t\over 2\pi} e^{-Q^2/4}-{1\over 4\pi}\int \frac {dX dq_y}{(2\pi)^2} 
e^{-q^2/2} \tilde {V}_o^Q(X,q_y)
\end{eqnarray}
where $\tilde {V}_o^Q(X,q_y)
=({\tilde V}_A+{\tilde V}_E \cos QX)/2$.
Using Eq.(\ref{eq:stiff}), we obtain the Hartree-Fock estimate of
isospin stiffness 
\begin{equation}
\rho_s^0={1\over 32\pi^2}\int_{0}^{\infty} dq {\tilde V}_E(q) 
e^{-q^2/2} q^3.
\end{equation}
Now we want to calculate the fluctuation corrections to $\rho_s^0$  
using the Hartree-Fock-RPA scheme.
The Green function $G^{\sigma \sigma^\prime}_{X X^\prime}(\tau-\tau^\prime)$ 
is defined as follows
\begin{equation}
G^{\sigma \sigma^\prime}_{X X^\prime}(\tau-\tau^\prime)\equiv 
\langle {\rm T}_\tau C_{\sigma X}(\tau)
C_{\sigma^\prime X^\prime}^\dagger(\tau^\prime)\rangle. 
\end{equation}
The retarded Green function $G^{\sigma \sigma^\prime}_{X X^\prime}(\omega)$
of the Hartree-Fock mean field Hamiltonian is given by 
\begin{equation}
G^{\sigma \sigma^\prime}_{X X^\prime}(\omega)=\delta_{X,X^\prime}\,
{1\over 2}\;e^{{i\over 2} (\sigma-\sigma^\prime)Q X} \Biggl\{ \frac {1}
{\omega-\epsilon_\alpha-i\eta} + \sigma \sigma^\prime 
\frac {1} {\omega-\epsilon_\beta+i\eta} \Biggr\} 
\end{equation}
where $\epsilon_\alpha=-t e^{-Q^2/4}-\sum_X \langle 0 X|\tilde {V}_0^Q|X 0
\rangle$, 
$\epsilon_\beta=t e^{-Q^2/4}-\sum_X \langle 0 X|\tilde {V}_x^Q|X 0\rangle$, 
and $\tilde {V}_x^Q(X,q_y)=(\tilde {V}_A-\tilde {V}_E \cos QX)/2$.
$\delta_{X,X^\prime}$ indicates that the system is
translationally invariant along the ${\hat y}$-direction in our chosen gauge 
and we only keep a single index $X$ from now on.
The RPA ground state energy can be calculated as follows\cite{Negele} 
\begin{equation}
E_{\rm RPA}=\sum_{n=2}^\infty {i\over 2n} 
\langle\Phi|(D\, V)^n|\Phi\rangle
\end{equation}
where $D$ is the Hartree-Fock particle-hole propagator defined by
\begin{eqnarray}
D^{\sigma \sigma^\prime}_{X X^\prime}(\omega)&=&\int \frac 
{d\omega^\prime}{2\pi}
G^{\sigma \sigma^\prime}_{X}(\omega+\omega^\prime)
G^{\sigma^\prime \sigma}_{X^\prime}
(\omega^\prime) \nonumber\\
&=&{i\over 4}\; \sigma \sigma^\prime e^{{i\over 2} 
(\sigma-\sigma^\prime)Q(X-X^\prime)} A(\omega) 
\end{eqnarray}
where $\Delta_{\rm x}(Q)=\Delta_{\rm SAS}\,e^{-Q^2/4}
+\sum_X \langle 0 X|\tilde {V}_E \,
\cos QX|X 0\rangle$ represents the $Q$-dependent exchange-enhanced gap  
with $\Delta_{\rm SAS}=2t$ and $A(\omega)=(\frac {1} {\omega-\Delta_{\rm x}(Q)
+i\eta}-\frac {1} {\omega+\Delta_{\rm x}(Q)-i\eta})$.
The strong enhancement of the tunneling gap due to the Coulomb exchange 
energy stabilizes our perturbative analysis. 
When $\sigma$ and $\sigma^\prime$ are opposite to each other,
$D^{\sigma \sigma^\prime}_{X X^\prime}$ 
represents an interlayer tunneling process which picks up  
an Aharonov-Bohm phase of 
$e^{\pm i Q(X-X^\prime)}$. 
We first consider the {\em n-th}
order bubble diagram (refer to Fig.(1))
\begin{eqnarray}
E^{(n)}_{\rm RPA} &=& \frac {(-i)^{n-1}}{2n} \sum_{\{\sigma_i,X_i\}}
\int \frac {d\omega}{2\pi}\,
D^{\sigma_{2n} \sigma_1}_{X_1 X_2}(\omega)\cdots              
D^{\sigma_{2n-2} \sigma_{2n-1}}_{X_{2n-1} X_{2n}}(\omega) \nonumber\\ 
& &\langle X_1 X_4 | V_{\sigma_1 \sigma_2} | X_2 X_3\rangle \cdots 
\langle X_{2n-1} X_{2} | V_{\sigma_{2n-1} \sigma_{2n}} 
| X_{2n} X_1 \rangle.
\end{eqnarray}
We notice that the momentum transfer $X=X_{2i-1}-X_{2i}$ is conserved.
The sum over isospins can be performed using the following trick.
Since $X$ stays the same for all the bubbles, the phase factor in 
$D^{\sigma_{2n} \sigma_1}_{X_1 X_2}$ can be rearranged
as follows
\begin{equation}
D^{\sigma_{2n} \sigma_1}_{X_1 X_2}(\omega)\cdots 
D^{\sigma_{2n-2} \sigma_{2n-1}}_{X_{2n-1} X_{2n}}(\omega)
=\left[i A(\omega)\right]^n \prod_{i=1}^{n}
\biggl[\frac {\sigma_{2i-1}\sigma_{2i}}{4} e^{-{i\over 2}
(\sigma_{2i-1}-\sigma_{2i}) Q X}\biggr] .
\label{Phase}
\end{equation}
Using Eq.(\ref{Phase}), the isospin-sum can 
be factorized and we 
obtain the following useful relations   
\begin{equation}
{1\over 4}\sum_{\sigma_{1},\sigma_{2}}
\sigma_{1}\sigma_{2} e^{-{i\over 2} (\sigma_{1}-\sigma_{2}) Q X}
\langle X_1 X_4 | V_{\sigma_1 \sigma_2} | X_2 X_3 \rangle
=V_x^Q(X,X_1-X_3) .
\end{equation}
Since our gauge choice conserves the momentum along 
the ${\hat y}$-direction, 
we finally obtain   
\begin{equation}
\frac {E^{(n)}_{\rm RPA}}{A}={i\over 2n} {1\over L^2}\sum_{X,q_y}
\int \frac {d\omega}{2\pi}\,
\biggl[{1\over 2\pi} e^{-q^2/2}
\tilde {V}_x^Q(X,q_y) A(\omega)\biggr]^n .
\end{equation}
The sum of $E^{(n)}_{\rm RPA}$ with respect to $n$ yields 
\begin{equation}
\frac {E_{\rm RPA}}{A}={i\over 2}\,{1\over L^2}\sum_{X,q_y}\, \int \frac 
{d\omega}{2\pi} 
\biggl\{-\ln \left[1-{1\over 2\pi} e^{-q^2/2}
\tilde {V}_x^Q(X,q_y) A(\omega) \right] \nonumber\\
-{1\over 2\pi} e^{-q^2/2}  
\tilde {V}_x^Q(X,q_y) A(\omega) \biggr\} .
\end{equation}
The integration over frequency $\omega$ can be explicitly 
performed and $E_{\rm RPA}$ can be written by  
\begin{eqnarray}
\frac {E_{\rm RPA}}{A}&=&{1\over 2 L^2}\sum_{X,q_y}\,
\biggl\{\sqrt{\Delta_{\rm x}^2(Q) + {1\over \pi} 
e^{-q^2/2} \tilde {V}_x^Q(X,q_y) \Delta_{\rm x}(Q)}
-\Delta_{\rm x}(Q)\biggr\} \nonumber\\ 
& &-{1\over 2 L^2}\sum_{X,q_y}\,{1\over 2\pi} 
e^{-q^2/2} \tilde {V}_x^Q(X,q_y).
\label{E_RPA}
\end{eqnarray}
The last term in Eq.(\ref{E_RPA}) plus $E_{HF}$ does not have a $Q$-dependence.
By expanding both $\Delta_{\rm x}(Q)$ and $\tilde {V}_x^Q$
upto the quadratic order in $Q$, we obtain  
\begin{eqnarray}
\rho_s&=&\int \frac {d^2 q}{(2\pi)^2} \Biggl\{  
\frac {{1\over 16\pi} e^{-q^2/2} q^2 \tilde {V}_E\, \Delta_{\rm x0}-
2 e^{-q^2/2} \rho_s^0 \tilde {V}_x}
{\sqrt{\Delta_{\rm x0}^2+{1\over \pi} e^{-q^2/2} \tilde {V}_x 
\Delta_{\rm x0}}} \nonumber\\
&+&\frac {
4\pi \rho_s^0 \left(\sqrt{\Delta_{\rm x0}^2+{1\over \pi} e^{-q^2/2}
\tilde {V}_x \Delta_{\rm x0}}-\Delta_{\rm x0}\right)}
{\sqrt{\Delta_{\rm x0}^2+{1\over \pi} e^{-q^2/2} \tilde {V}_x
\Delta_{\rm x0}}}\Biggr\}  
\end{eqnarray}
where $\tilde {V}_x=(\tilde {V}_A-\tilde {V}_E)/2$ and 
$\Delta_{\rm x0}=\Delta_{\rm x}(Q=0)$.

It is well-known that the coherent double-layer quantum Hall state is 
unstable due to the presence of a charge density wave(CDW) instability 
at a finite wave-vector $k\ell\cong 1$ for 
$d/\ell\geq 1.2$\cite{Fertig,MacdPlatBoe,FertigDB}.
In order to correctly capture this important quantum fluctuations,
we take into account the vertex corrections, {\em i.e.}, a sum of ladders
in the polarization insertions in Fig.(1).
It amounts to replacing $\Delta_{\rm x}(Q)$ with the correct dispersion
relation $\omega({\bf q},Q)$ of the collective mode in Eq.(\ref{E_RPA})
\begin{equation}
\omega({\bf q},Q)=\sqrt{D_z({\bf q},Q) D_y({\bf q},Q)}
\end{equation}
where $D_z({\bf q},Q)$ and $D_y({\bf q},Q)$ are given by 
\begin{equation}
D_z({\bf q},Q)=\Delta_{\rm SAS} + {1\over \pi} \tilde {V}^Q_x e^{-q^2/2} 
+\int {d^2 k\over (2\pi)^2} \tilde {V}_E(k)\cos Qk_y e^{-k^2/2}
-\int {d^2 k\over (2\pi)^2} \tilde {V}_A(k) e^{-k^2/2} e^{i{\bf k}
\times {\bf q}\cdot {\hat z}}
\end{equation}
\begin{equation}
D_y({\bf q},Q)=\Delta_{\rm SAS} 
+\int {d^2 k\over (2\pi)^2} \tilde {V}_E(k)\cos Qk_y e^{-k^2/2}
-\int {d^2 k\over (2\pi)^2} \tilde {V}_E(k)\cos Qk_y e^{-k^2/2} e^{i{\bf k}
\times {\bf q}\cdot {\hat z}}.
\end{equation}
Fig.(2) demonstrates that $\rho_s$ obtained by RPA with  
vertex corrections indeed exhibits    
a dramatic reduction from the Hartree-Fock {\em or} RPA
estimates. 
One can also notice that $\rho_s$ vanishes at $d/\ell\cong 1$
due to the proximity to the strong CDW instability. 
Quantum fluctuations can also reduce the normalized magnetization 
$<m_x>$---the {\em
order parameter} of the DLQHS. One can calculate $<m_x>$ using the following 
procedure\cite{DBPRBI}  
\begin{eqnarray}
<m_x> &=& - \lim_{t\rightarrow 0} \lim_{Q\rightarrow 0}
2\pi \frac {d\epsilon (Q,t)}{dt} \nonumber\\ 
&=&1-\int \frac {d^2 q}{2\pi} \Biggl\{
\frac {\Delta_{\rm x0} + {1\over 2\pi} e^{-q^2/2} \tilde {V}_x}
{\sqrt{\Delta_{\rm x0}^2+{1\over \pi} e^{-q^2/2} \tilde {V}_x
\Delta_{\rm x0}}} -1\Biggr\} . 
\end{eqnarray}
The renormalization of $<m_x>$ implies that 
the effective tunneling amplitude $t_R$ is  
reduced by a factor of $<m_x>$: $t_R=t<m_x>$.
The inset of Fig.(2) shows that $<m_x>$ via RPA with vertex 
corrections is strongly reduced from the Hartree-Fock {\em or}
RPA estimates. Based on the general field-theoretical point of view, 
we speculate that in the absence of tunneling, $<m_x>$ will vanish at
the same value of $d$ as $\rho_s$\cite{DBPRBI}. Here $<m_x>$ vanishes at 
$d/\ell\sim 0.7$, which is comparable to $d/\ell\sim 1$ from 
$\rho_s$.

The activation energy of the lowest charge excitations can be calculated 
from $\rho_s$ and $t_R$ obtained above. 
The subscript $R$ is omitted from now on. 
When the tunneling is very weak, {\em i.e.}, ${\tilde t}\leq 2.4\times 10^3\, 
{\tilde \rho}_s^3$, the lowest charge excitations are meron-pairs,
where ${\tilde t},{\tilde \rho}_s$ are measured in units of $e^2/\epsilon\ell$
\cite{DBPRBI,YangMacd}.
Since the merons carry charge $\pm {1 \over 2}e$ depending on the vorticity
and the core-spin configurations,
the energy $E_{MP}$ of meron pairs with charge $\pm e$ can be determined
by balancing the
Coulomb repulsion and the logarithmic attraction: 
$E_{MP}\sim 4\pi\rho_s$\cite{DBPRBI}. 
As $t$ increases {\em or} $\rho_s$ decreases, it becomes too costly to 
flip the isospins over a broad area due to the large tunneling energy cost. 
Hence the meron-pairs are confined by a linear string tension 
$T_0=8(t\rho_s/2\pi)^{1/2}$  
and the lowest charge excitations become domain-wall string 
solitons(DWS)\cite{DBPRBI,Nick}.
The activation energy of the DWS can be estimated by 
balancing the Coulomb repulsion and linear string tension:  
$E_{DWS}\sim (e^2 T_0/\epsilon)^{1/2}$\cite{DBPRBI,Nick}. 
We have chosen $t/(e^2/\epsilon\ell)\cong 4\times 10^{-3}$\cite{comment1}.
Fig.(3) shows the $d/\ell$-dependence of the activation energy. 
Since one needs to create a charge {\em neutral} excitation at $\nu_T=1$, 
$E_A$ is multiplied by a factor of two.
Below $d/\ell\sim 0.7$, the meron pairs are the lowest charge
excitations and above are the DWS.
The theoretical result thus obtained demonstrates a nice qualitative 
agreement with the experiment\cite{Hyndman}. 
When $d/\ell$ vanishes, one recovers a skyrmion-antiskyrmion 
pair energy $8\pi\rho_s$\cite{Sondhi}. 
In real experimental systems, when the layer separation becomes small, 
the interlayer tunneling becomes much easier and 
the system acts like a single wide quantum well. 
As the Zeeman gap is much smaller than $\Delta_{\rm SAS}$,
the real spin degrees of freedom will become important and
the lowest charge excitations are {\em real spin}-textured   
quasiparticles. 

To summarize, we have calculated the renormalized isospin 
stiffness $\rho_s$ and
the effective tunneling amplitude $t_R$ within the Hartree-Fock-RPA formalism.
We have explicitly shown that $\rho_s$ vanishes 
at $d/\ell\cong 1$.
We have also estimated the $d/\ell$-dependent activation energy of 
the topological excitations. 

\acknowledgements{
It is my pleasure to acknowledge useful conversations with S.M. Girvin, 
A.H. MacDonald, K. Mullen, and 
S.Q. Murphy. The work was supported by NSF DMR-9502555.
I want to acknowledge the Aspen center for physics where part of this work 
has been performed.
}

\begin{figure}
\caption{The {\em n-th} order RPA bubble diagram.}
\label{fig1}
\end{figure}

\begin{figure}
\caption{
The renormalized isospin stiffness $\rho_s$: the solid line 
stands for the Hartree-Fock result $\rho_s^0$, the dashed line
for the RPA, and the long-dashed line for the RPA with 
the vertex corrections.
The inset shows the corresponding renormalized magnetizations $<m_x>$.
}
\label{fig2}
\end{figure}

\begin{figure}
\caption{The activation energy of the lowest charge excitations.}
\label{fig3}
\end{figure}


\begin{references}

\bibitem{GirvinMacdBook} S.M. Girvin and A.H. MacDonald, to appear 
in {\em Novel Quantum Liquids in 
Low-Dimensional Semiconductor Structures}, edited by S. Das Sarma and
A. Pinczuk (Wiley, New York, 1996).

\bibitem{DBPRL} Kun Yang, K. Moon, L. Zheng, A.H. MacDonald, 
S.M. Girvin, D. Yoshioka, and Shou-Cheng Zhang,
{\em Phys. Rev. Lett.} {\bf 72}, 732 (1994).

\bibitem{Sheena} S.Q. Murphy, J.P. Eisenstein, G.S. Boebinger, 
L.N. Pfeiffer, and K.W. West,  
{\em Phys. Rev. Lett.} {\bf 72}, 728 (1994).

\bibitem{WenandZee} X.G. Wen and A. Zee, 
{\em Phys. Rev. Lett.} {\bf 69}, 1811 (1992).

\bibitem{Ezawa} Z.F. Ezawa and A. Iwazaki, {\em Int. J. of Mod. Phys. B}
{\bf 19}, 3205 (1992).

\bibitem{DBPRBI} K. Moon {\em et al.}, {\em Phys. Rev. B} {\bf 51}, 
5138 (1995); Kun Yang {\em et al.}, {\em Phys. Rev. B} {\bf 54}, 11644 (1996).

\bibitem{Fertig} H.A. Fertig, {\em Phys. Rev. B} {\bf 40}, 1087 (1989).

\bibitem{Shayegan} T.S. Lay {\em et al.}, 
{\em Phys. Rev. B} {\bf 50}, 17725 (1994).

\bibitem{MacdPlatBoe} A.H. MacDonald, P.M. Platzman, and G.S. Boebinger,
{\em Phys. Rev. Lett.} {\bf 65}, 775 (1990).

\bibitem{Hyndman} R.J. Hyndman {\em et al.},  
{\em preprint} (Univ. of Nottingham). 

\bibitem{TLho} Tin-Lun Ho,  
{\em Phys. Rev. Lett.} {\bf 73}, 874 (1994). 

\bibitem{FertigDB} R. C\^ot\'e, L. Brey, H.A. Fertig, and A.H. MacDonald,  
{\em Phys. Rev. B} {\bf 51}, 13475 (1995); R. C\^ot\'e and A.H. MacDonald, 
{\em Phys. Rev. B} {\bf 44}, 8759 (1991).

\bibitem{YangMacd} K. Yang and A.H. MacDonald,  
{\em Phys. Rev. B} {\bf 51}, 17247 (1995).

\bibitem{Negele} J.~W. Negele and H. Orland, {\em Quantum Many-Particle 
Systems} (Addison-Wesley 1987). 

\bibitem{bak} Per Bak, {\em Rep. Prog. Phys.} {\bf 45}, 587 (1982).

\bibitem{Nick} N. Read,   
{\em Phys. Rev. B} {\bf 52}, 1926 (1995). 

\bibitem{comment1} We notice that $<m_x>$ has {\em much} stronger dependence on
the weak tunneling than $\rho_s$. 

\bibitem{Sondhi} S.L. Sondhi, A. Karlhede, S.A. Kivelson, and E.H. Rezayi,  
{\em Phys. Rev. B} {\bf 47}, 16419 (1993). 

\end{references}
\end{document}